\documentclass[twocolumn, prl]{revtex4}

\usepackage{amsmath}
\usepackage{amssymb}
\usepackage{dsfont}
\usepackage{color}
\usepackage{hyperref}
\usepackage{graphicx}
\usepackage{epstopdf}
\usepackage{soul}
\usepackage{color}
\usepackage{braket}
\usepackage[normalem]{ulem}
\usepackage{xcolor}
\hypersetup{
  colorlinks   = true, 
  urlcolor     = blue, 
  linkcolor    = blue, 
  citecolor   = blue 
}

\usepackage{textcomp}

\DeclareTextCommandDefault{\textcopyright}{\textcircled{c}}

\begin{document}
\title{Observation of robust flat-band localization in driven photonic rhombic lattices}
\author{Sebabrata Mukherjee}
\email[Corresponding author: ]{s.mukherjee@hw.ac.uk}
\affiliation{Institute of Photonics and Quantum Sciences, School of Engineering $\&$ Physical Sciences, Heriot-Watt University, Edinburgh, EH14 4AS, United Kingdom}
\author{Robert R. Thomson}
\affiliation{Institute of Photonics and Quantum Sciences, School of Engineering $\&$ Physical Sciences, Heriot-Watt University, Edinburgh, EH14 4AS, United Kingdom}

\begin{abstract}
We demonstrate that a flat-band state in a quasi-one-dimensional rhombic lattice is robust in the presence of external drivings along the lattice axis. The lattice was formed by periodic arrays of evanescently coupled optical waveguides, and the external drivings were realized by modulating the paths of the waveguides. We excited a superposition of flat-band eigenmodes at the input and observed that this state does not diffract in the presence of static as well as high-frequency sinusoidal drivings. This robust localization is due to destructive interference of the analogous wavefunction and is associated with the symmetry in the lattice geometry. We then excited the dispersive bands and observed Bloch oscillations and coherent destruction of tunneling.
{\textcopyright} 2017 Optical Society of America. \\

\smallskip
\noindent \textbf{OCIS codes:} (130.0130) Integrated optics; (000.1600) Classical and quantum physics; (130.2790) Guidedwaves;(230.7370) Waveguides; (350.3950) Micro-optics.
\end{abstract}

\maketitle


Nearly a century ago, it was predicted that an electron in a periodic potential and uniform electric field can exhibit periodic Bloch oscillations~\cite{bloch1929quantenmechanik, fukuyama1973tightly, dahan1996bloch, peschel1998optical, morandotti1999experimental, chiodo2006imaging} 
due to the formation of Wannier-Stark ladder. 
An ac electric field, on the other hand, can renormalize the effective tunneling probability~\cite{dunlap1986dynamic} causing interesting localization effects such as dynamic localization~\cite{dreisow2008spectral}.
The dynamics of electrons is also influenced by external magnetic flux, the presence of disorder and particle interactions. It is of great interest to study these fundamental transport phenomena in a lattice geometry supporting non-dispersive (flat) bands; see ref.~\cite{tasaki2008hubbard, huber2010bose, nictua2013spectral,  guzman2014experimental, mukherjee2015observation, vicencio2015Observation, taie2015coherent, xia2016demonstration, mukherjee2015rhombic, molina2015flat, nandy2016engineering, diebel2016conical, 
weimann2016transport, Yang2016, Owen2016} for recent works on flat-band lattices. 
Flat-band eigenmodes are degenerate with infinite effective mass; i.e.~a flat-band state has no time evolution.
However, in the presence of external electric and magnetic fields, the initial flat-band state of the instantaneous Hamiltonian, $\mathcal{\hat H}(t\!=\!0)$, can exhibit complicated dynamics, such as Bloch oscillations associated with Landau-Zener tunneling~\cite{cai2013quantum, Khomeriki2016landau}, depending on the lattice geometry and the external drivings. 
These intriguing phenomena are at the heart of condensed matter physics, and can be investigated in the system of ultracold atoms in optical lattices~\cite{bloch2012quantum} and periodic arrays of coupled optical waveguide (photonic lattices)~\cite{longhi2009quantum, garanovich2012light}.
In fact, localized flat-band state was experimentally demonstrated using lattice geometries such as Lieb~\cite{mukherjee2015observation, vicencio2015Observation, taie2015coherent, xia2016demonstration}, Kagome~\cite{zong2016observation}, rhombic~\cite{mukherjee2015rhombic} and stub~\cite{baboux2016bosonic}.

\begin{figure}[t]
\centering
\includegraphics[width=8.6 cm]{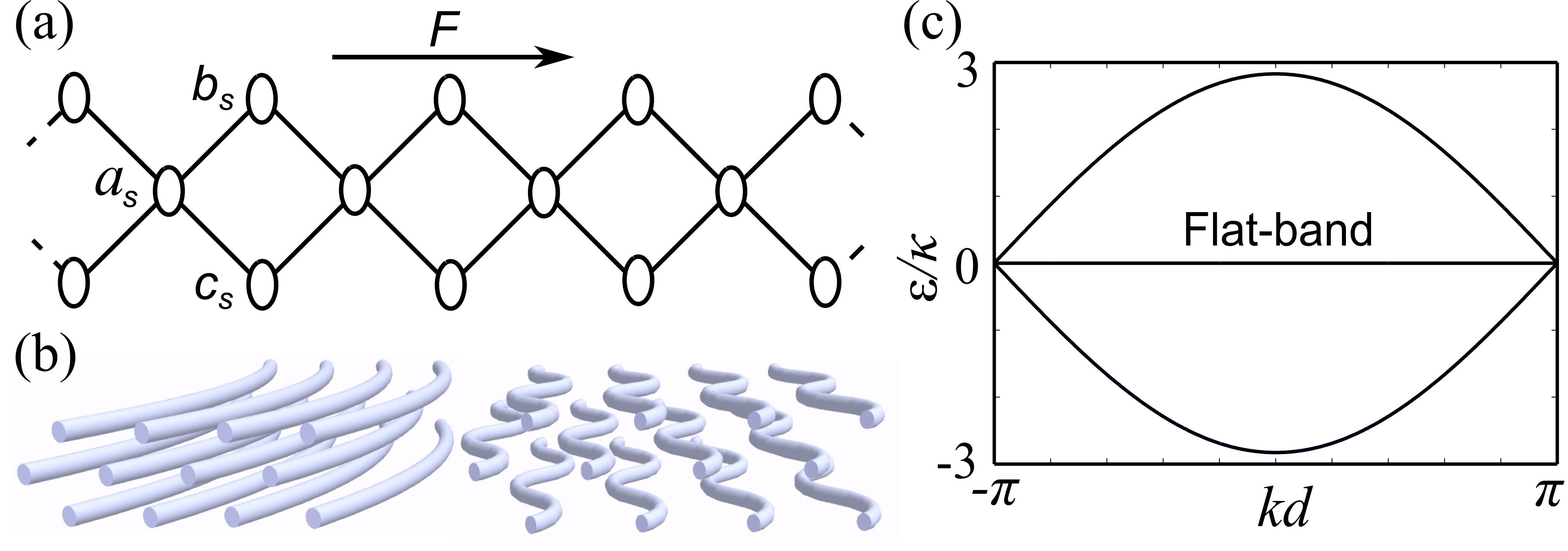}
\caption{(a) A quasi-one-dimensional rhombic lattice driven by an external force, $F$. The unit cell contains three sites, $a$, $b$ and $c$. (b) 
In the photonic setup, the static and sinusoidal drivings are implemented by modulating the paths of the waveguides; see text. (c) Band structure calculated by diagonalizing the Fourier transformed Hamiltonian with $F\!=\!0$. In this situation, the non-dispersive (flat) band can be excited by initially exciting $\{E^b_s, E^c_s\}\!=\!\{1/\sqrt{2}, -1/\sqrt{2}\}$ state, see ref.~\cite{mukherjee2015rhombic}.}
\label{Fig-1}
\end{figure}

In this work, we consider a quasi-one-dimensional photonic rhombic [or diamond, Fig.~\ref{Fig-1}~(a, b)] lattice with three sites ($a$, $b$ and $c$) per unit cell. This lattice  geometry has been previously used to theoretically study various interesting effects such as magnetic field induced Aharonov-Bohm caging for both interacting and non-interacting particles~\cite{vidal2000interaction, creffield2010coherent, longhi2014aharonov}, Anderson localization~\cite{leykam2013flat}, conservative and PT-symmetric compactons~\cite{yulin2013conservative} and Landau-Zener Bloch oscillations~\cite{Khomeriki2016landau}.
In the nearest neighbor tight binding approximation, diagonalizing the Fourier-transformed single particle Hamiltonian of the rhombic chain, one obtains the following dispersion relations~\cite{vidal2000interaction}: $\varepsilon^{0,\pm}\!=\!0,\pm2\kappa \sqrt{1+\cos(kd)}$, where $\kappa$ is the nearest neighbor hopping amplitude (or coupling constant), $d$ is the lattice constant and  $k$ is the quasimomentum. The lattice supports three energy bands, two dispersive ($\varepsilon^{\pm}$) and one non-dispersive ($\varepsilon^{0}$) in the middle; see Fig.~\ref{Fig-1}(c). A flat-band state can be excited by initially exciting the $b$ and $c$ sites of a unit cell with equal intensity and opposite phases~\cite{mukherjee2015rhombic}.

Under the influence of a {\it weak} analogous static force, we show that a state prepared in the dispersive bands of the rhombic lattice, exhibits Bloch oscillations~\cite{dahan1996bloch, peschel1998optical, morandotti1999experimental, chiodo2006imaging}, however, a flat-band state keeps the compactness of the initial state (as~$\partial_t \psi_k^0(t)\!=\!0$; see also R.~Khomeriki et al.~\cite{Khomeriki2016landau}).
We then show that an analogous high-frequency sinusoidal force modifies the band structure keeping the flat-band unaffected. Interestingly, the initial flat-band state overlaps only with the degenerate Floquet eigenstates, and hence, remains localized. For a state in the dispersive bands, on the other hand, we observed renormalization of tunneling~\cite{holthaus1993quantum, grifoni1998driven, della2007visualization, mukherjee2016experimental}, as would be expected. 

\begin{figure}[t]
\centering
\includegraphics[width=8.6 cm]{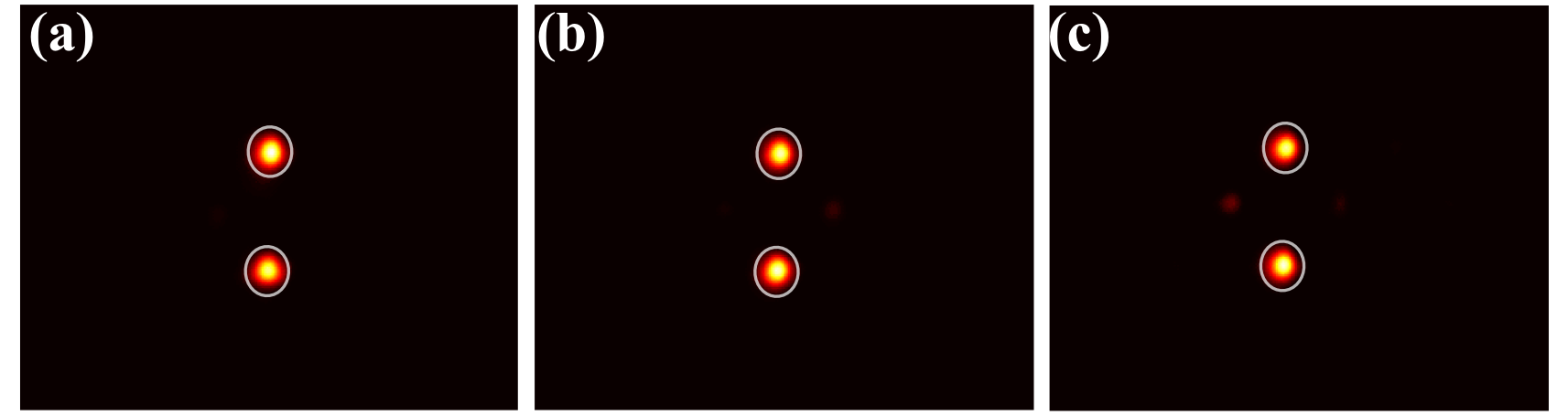}
\caption{Robust flat-band localization in the presence of static driving. (a-c) Measured output intensity distributions after a propagation of $10$, $30$ and $70$~mm respectively. The flat-band was excited by launching $\{E^b_6, E^c_6\}\!=\!\{1/\sqrt{2}, -1/\sqrt{2}\}$ state at the input of circularly curved photonic rhombic lattices. Each image is normalized such that the total output power is $1$. The circles indicate the initially (at $z\!=\!0$) excited lattice sites.}
\label{Fig-2}
\end{figure}
The propagation of light across a one-dimensional photonic lattice where the axes of all the waveguides are simultaneously and slowly bending along the lattice axis ($x$), is governed by the following scalar-paraxial equation~\cite{longhi2009quantum, garanovich2012light}:
\begin{eqnarray}
i\lambdabar \frac {\partial}{\partial z}E(x, z)\!=\!\Big[\!-\frac{\lambdabar^2}{2n_0} \frac {\partial ^2}{\partial x^2}\!-\!\Delta n(x) \!+\!F(z)x \Big] E(x, z) \label{1}
\end{eqnarray}
where $E(x, z)$ is the electric field envelope of the light waves, $n_0$ is the average refractive index of the medium, $\lambda=2\pi\lambdabar$ is the free-space wavelength, $\Delta n(x)$ is the transverse refractive index profile, $z$ plays the role of time and $F\!=\!-n_0\partial_z^2 x_0(z)$. Here $x_0(z)$ describes the bending profile of each waveguide and $(x, z)$ is the coordinate system moving with the lattice. Equation~(\ref{1}) is analogous to the Schr\"odinger equation for a quasiparticle (e.g.~an electron) in a one-dimensional lattice driven by an external force, $F$.
For weak external fields, neglecting the excitation of higher Bloch bands, the following coupled-mode equations are obtained~\cite{longhi2014aharonov, Khomeriki2016landau} for a rhombic lattice
\begin{align}
&\big[i \partial_z+2s\beta \big]E^a_s\!=\!-\kappa \big(E^b_s+E^b_{s-1}+E^c_s+E^c_{s-1}\big) \nonumber \\
&\big[i \partial_z+(2s+1)\beta \big]E^b_s\!=\!-\kappa \big(E^a_s+E^a_{s+1} \big) \nonumber\\
&\big[i \partial_z+(2s+1)\beta \big]E^c_s\!=\!-\kappa \big(E^a_s+E^a_{s+1} \big) \label{2}
\end{align}
where $\{E^a_s, E^b_s, E^c_s\}$ are the electric field amplitudes of the light waves at the $a$, $b$ and $c$ sites of the $s$-th unit cell respectively. The analogous static electric field is realized by circularly curving the waveguide axis, $x_0^2\!=\!R^2\!-\!z^2$ where $R$ is the radius of curvature, implying almost a linear shift in propagation constant (site energy) along the lattice axis, $\beta\!=\!n_0 d/(2R\lambdabar)$. Similarly, a sinusoidal ac field is realized by sinusoidally modulating the waveguide paths along the lattice axis. In this case $x_0\!=\!A\sin(\omega z)$ and $\beta(z)\!=\!K\sin(\omega z)$, where $A$ and $\omega$ are the amplitude and frequency of the modulation, and $K\!=\!n_0A\omega^2d/(2\lambdabar)$; see ref.~\cite{longhi2006observation, della2007visualization, dreisow2008spectral, szameit2009polychromatic, longhi2011tunneling, mukherjee2015modulation, mukherjee2016experimental} and the references therein for more details on driven photonic lattices. The intensity distribution at the output of the photonic lattice, for a given input excitation, is obtained by numerically solving Eq.~(\ref{2}).

Curved photonic lattices were fabricated inside a borosilicate substrate (Corning Eagle$^{2000}$) using ultrafast laser inscription technique~\cite{davis1996writing}. The substrate was mounted on $x$-$y$-$z$ translation stages and each waveguide was fabricated by translating the substrate through the focus of the fs laser pulses (350~fs, 500~kHz, 1030~nm). Fabrication parameters were optimized to realize well-confined single-mode waveguides for operation at $780$~nm; see ref.~\cite{mukherjee2015observation, mukherjee2015rhombic} for fabrication details. We note that carefully engineered photonic lattices were previously used to experimentally simulate various quantum phenomena including the investigation of localization effects, i.e.~the inhibition of transport, caused by disorders~\cite{schwartz2007transport, martin2011anderson}, external fields~\cite {longhi2006observation, della2007visualization, dreisow2008spectral, szameit2009polychromatic, longhi2011tunneling, mukherjee2015modulation, mukherjee2016experimental} and Kerr nonlinearity~\cite {szameit2006two}. 

\begin{figure}[t]
\centering
\includegraphics[width=8.6 cm]{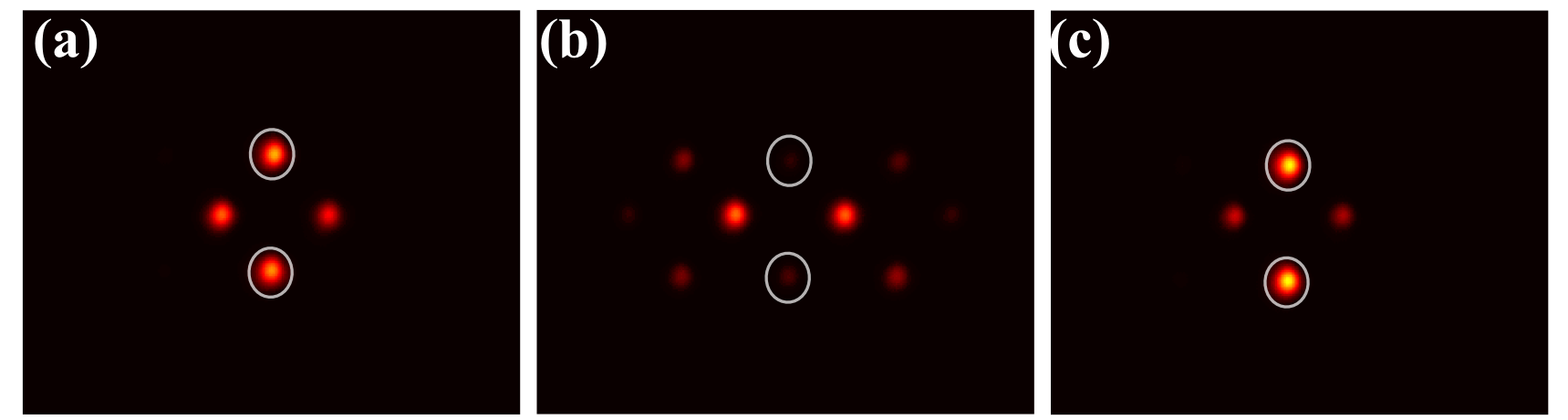}
\caption{Observation of Bloch oscillations when the dispersive bands are excited by launching the {\it equal phase state,} $\{E^b_6, E^c_6\}\!=\!\{1/\sqrt{2}, 1/\sqrt{2}\}$; see Fig.~\ref{Fig-4}. The propagation distances were $10$, $30$ and $70$~mm for (a), (b) and (c) respectively.}
\label{Fig-3}
\end{figure}
\begin{figure}[]
\centering
\includegraphics[width=8.6 cm]{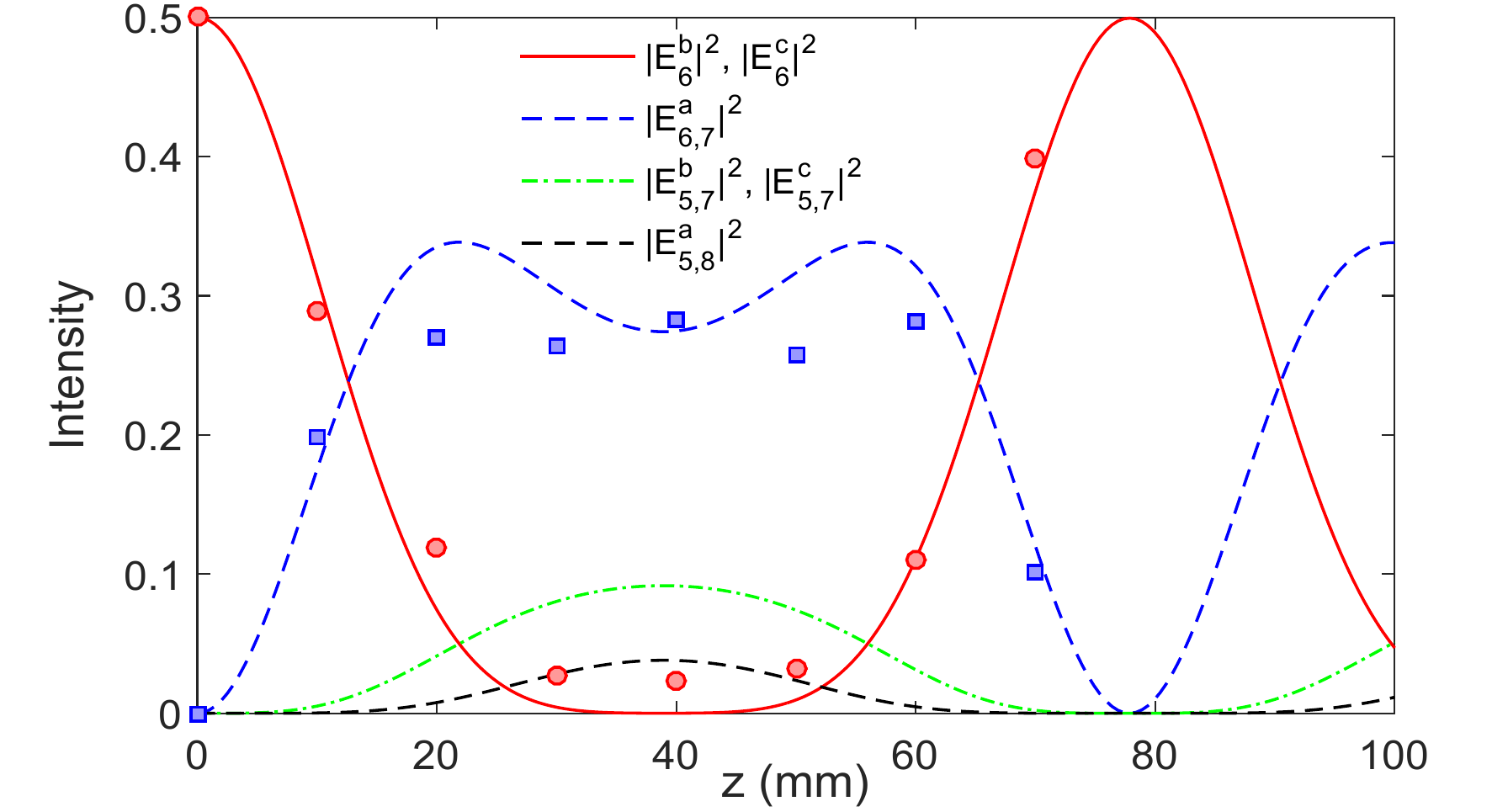}
\caption{Numerically calculated variation of light intensity along the propagation direction of a circularly curved photonic rhombic lattice for the initial condition, $\{E^b_6, E^c_6\}\!=\!\{1/\sqrt{2}, 1/\sqrt{2}\}$. Here $\kappa\!=\!0.034$~mm$^{-1}$ and $R\!=\!1.8$~m. Note that the period of the breathing motion of the 
intensity distribution is $z_B\!=\!2R\lambda/(n_0d)\!=\!77.87$~mm. The red circles and the blue squares are the measured values of the average intensities at the waveguides that were excited at the input [i.e.~$(|E^b_6|^2+|E^c_6|^2)/2$] 
and their nearest neighbor sites [i.e.~$(|E^a_6|^2+|E^a_7|^2)/2$)] respectively.}
\label{Fig-4}
\end{figure}
%

 
It is evident from Eq.~(\ref{2}) that an initial state occupying only $b_s$ and $c_s$ sites with equal intensity and opposite phases, which excites flat-band eigenmodes, does not have any time evolution in a static field. 
To demonstrate this, seven circularly curved finite photonic lattices (with propagation lengths 10~mm to 70~mm in steps of 10~mm) were fabricated with 12 unit cells and radius of curvature $R\!=\!1.8$~m. For all the lattices waveguide-to-waveguide separation was $d/\sqrt{2}\!=\!17$~$\mu$m, and the coupling constant was $\kappa\!=\!0.034$~mm$^{-1}$. First, we excited an equal superposition of flat-band eigenmodes by launching the input state $\{E^b_6, E^c_6\}\!=\!\{1/\sqrt{2}, -1/\sqrt{2}\}$ and measured the output intensity distributions at seven different values of propagation distances. As shown in Figs.~\ref{Fig-2}~(a)-(c), we observe that this input state remains localized without significantly tunneling to other lattice sites. However, when the dispersive bands were excited launching an orthogonal mode (we call it {\it equal phase mode,} $\{E^b_6, E^c_6\}\!=\!\{1/\sqrt{2}, 1/\sqrt{2}\}$), we observe a breathing motion [Fig.~\ref{Fig-3} and \ref{Fig-4} ] of the 
intensity distribution which is the characteristic of Bloch oscillations. In Fig.~\ref{Fig-4}, the measured variations of the light intensities (red circles: $(|E^b_6|^2+|E^c_6|^2)/2$; blue squares: $(|E^a_6|^2+|E^a_7|^2)/2$) are plotted as a function of $z$ which are in good agreement with the numerically calculated results. We note that the input states were prepared using a zero-order nulled diffractive optical element; see ref.~\cite{mukherjee2015rhombic}. 

\begin{figure}[!t]
\centering
\includegraphics[width=8.6 cm]{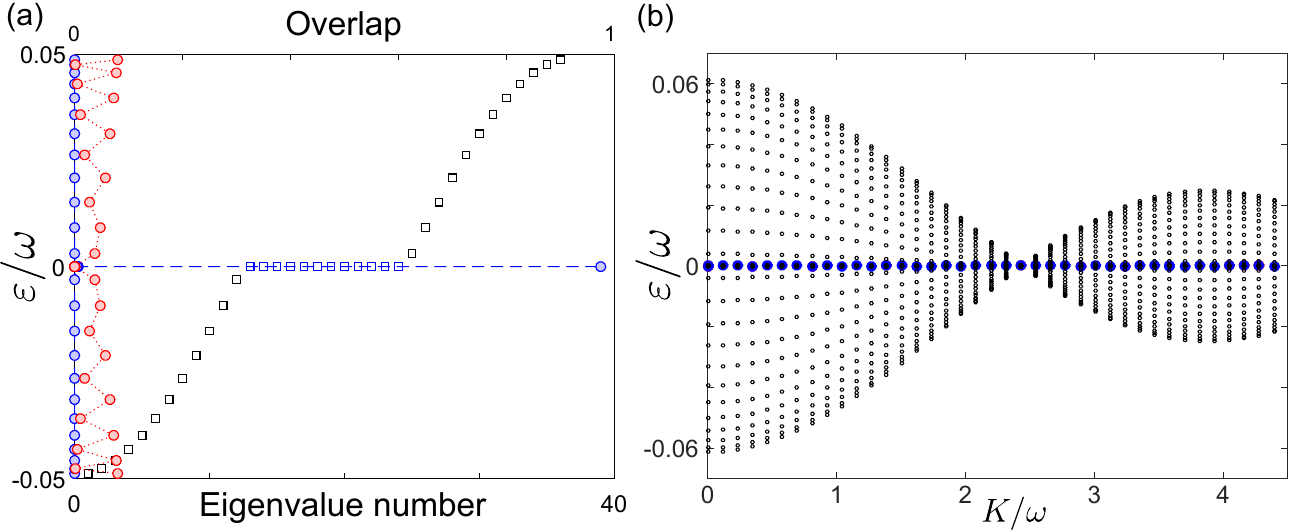}
\caption{(a) Numerically calculated quasienergy spectrum for a sinusoidally driven rhombic lattice with 12 unit cells (open squares). Here $\kappa\!=\!0.034$~mm$^{-1}$, $\omega/\kappa\!=\!46.2$ and $K/\omega\!=\!0.912$ ($A\!=\!4$~$\mu$m). Note that the driving does no destroy the flat-band, shown by the open blue squares. The blue (red) circles show the calculated values of the overlap of the initial flat-band state (equal phase state) with the Floquet eigenstates.
(b) Floquet spectrum of the as a function of $K/\omega$. When $\{E^b_6, E^c_6\}\!=\!\{1/\sqrt{2}, -1/\sqrt{2}\}$ state is excited, the eigenstates at zero quasienergies (shown by the blue circles) are only excited irrespective of the value of $K/\omega$. The dispersive bands collapse at $K/\omega\!=\!2.405$ causing coherent destruction of tunneling; see Fig.~\ref{Fig-6}(g).}
\label{Fig-5}
\end{figure}

In the presence of a sinusoidal driving, the Hamiltonian that describes the system is periodic in $z$, the analogous time, i.e.~$\mathcal{\hat H}(z)\!=\!\mathcal{\hat H}(z+z_0)$ where $z_0\!=\!2\pi/\omega$. In this situation, using Floquet theory~\cite{holthaus1993quantum} the quasienergy spectrum is obtained by diagonalizing the evolution operator defined as $\hat U\!=\!\mathcal{T} \exp[-i \int_0^{z_0} \mathcal {\hat H}(z') dz' ],$ where $\mathcal{T}$ indicates the time ordering. 
The Floquet quasienergy spectrum for a high-frequency sinusoidal driving (with $K/\omega\!=\!0.912$) is shown in Fig.~\ref{Fig-5}(a) [open squares]. There are two important aspects to be noted. First, the flat-band is not destroyed by the ac driving and second, an initial flat-band state (i.e.~$\{E^b_6, E^c_6\}\!=\!\{1/\sqrt{2}, -1/\sqrt{2}\}$) overlaps only with the degenerate Floquet eigenstates; see the blue circles. Therefore, this input state is expected to remain localized. Figure~\ref{Fig-5}(b) shows the Floquet spectrum as a function of $K/\omega$; irrespective of the value of the $K/\omega$, the initial flat-band state overlaps only with the eigenmodes at zero quasienergies (shown by the blue circles).
Note that the dispersive bands of the lattice (pseudo) collapse~\cite{holthaus1993quantum, della2007visualization} when $\mathcal{J}_0(K/\omega)\!=\!0$ i.e.~$K/\omega\!=\!2.405$ [$\mathcal{J}$ is the Bessel function of the first kind]. In other words, a sinusoidal driving renormalizes the effective coupling constant as $|\kappa_{\text{eff}}/\kappa|\!=\!|\mathcal{J}_0(K/\omega)|$.

\begin{figure}[t]
\centering
\includegraphics[width=8.6 cm]{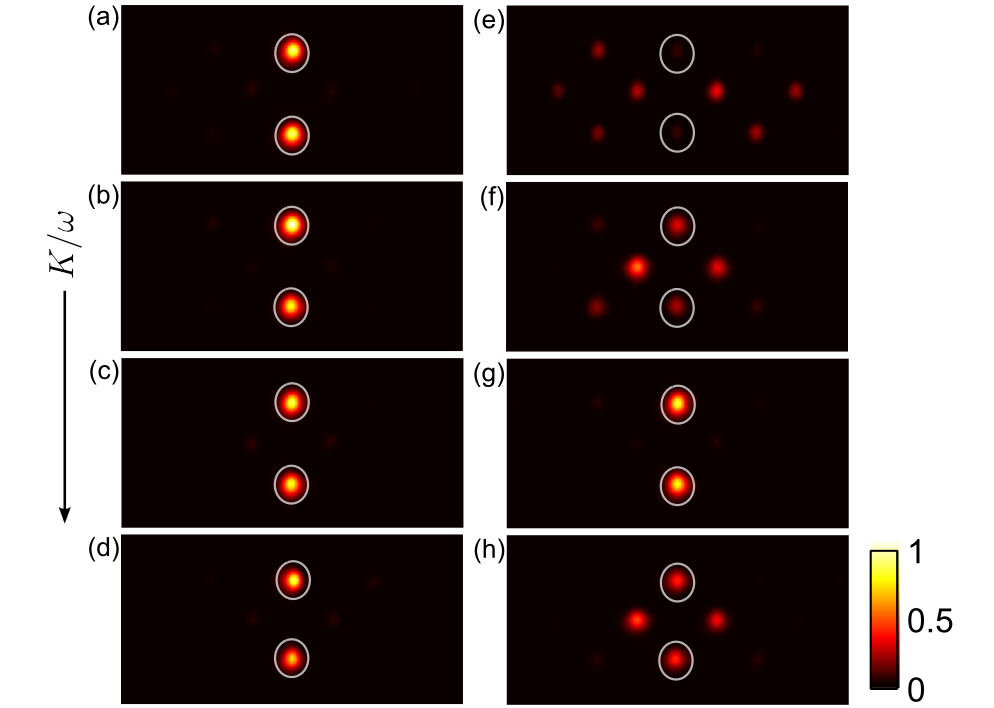}
\caption{Experimentally measured intensity distributions at the output of the 70-mm-long modulated lattices for four different values of amplitude of modulations ($A\!=\!6, 8, 10~\& ~12$~$\mu$m respectively). For the left column,  (a)-(d), the flat-band state was launched at the input. For the right column, (e)-(h), the dispersive bands were excited by launching the equal-phase state at the input. The flat-band state is robust in the presence of sinusoidal drivings. For the equal phase input state, the output intensity distribution is determined by $K/\omega$ as the effective coupling constant is renormalized as $|\kappa_{\text{eff}}/\kappa|\!=\!|\mathcal{J}_0(K/\omega)|$. Note that for (g), 
$K/\omega$ is close to the first root of $\mathcal{J}_0(K/\omega)$ causing coherent destruction of tunneling.}
\label{Fig-6} 
\end{figure}

\begin{figure}[t]
\centering
\includegraphics[width=8.6 cm]{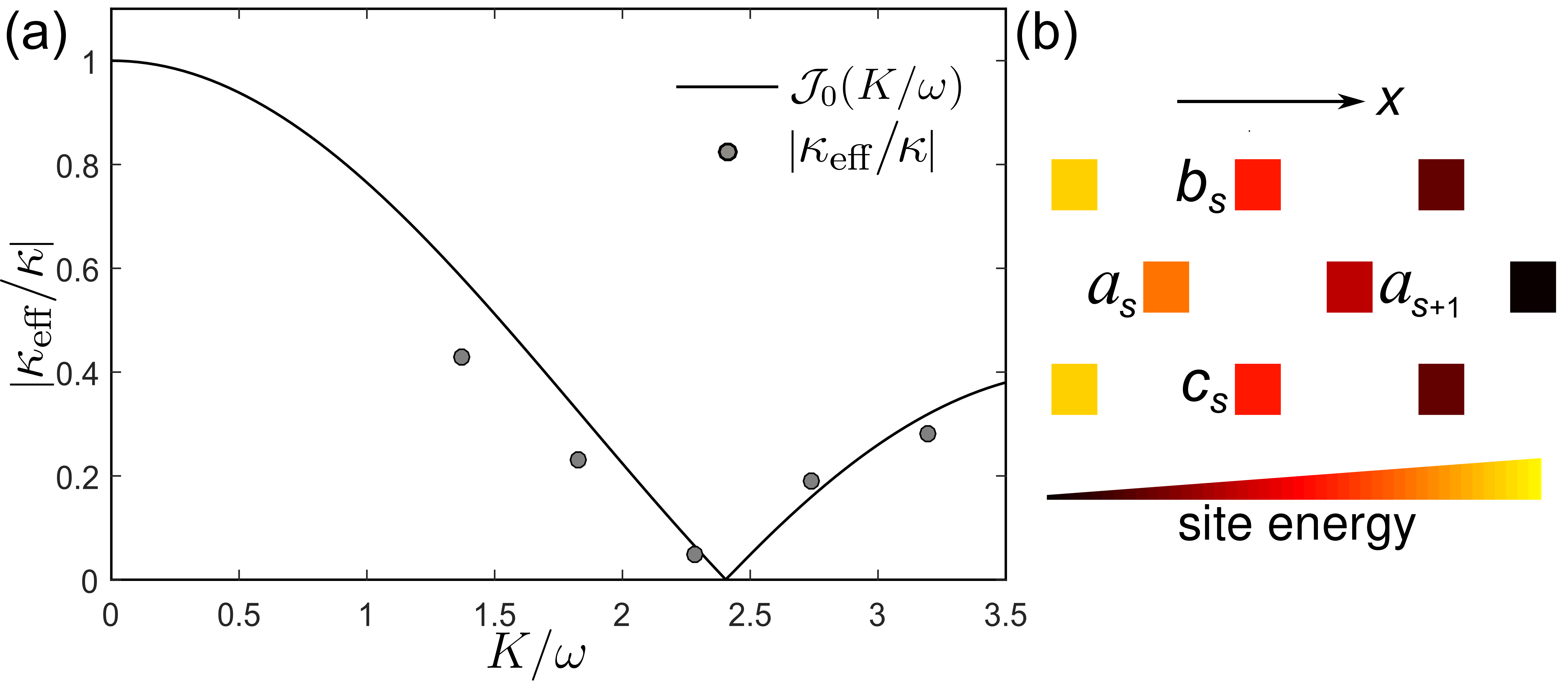}
\caption{(a) High-frequency sinusoidal driving renormalizes the effective coupling constant as shown by the solid line. The filled circles indicate the values of $|\kappa_{\text{eff}}/\kappa|$, estimated from the measured intensity distributions. (b) Analogous site energy of the lattice sites at a given $z$, in the presence of an external field along $x$-axis. The linear ramp of the site energy is determined by the strength of the external field and its direction. An initial flat-band state remains localized as the $b_s$ and $c_s$ sites have the same site energy which is shifted by $\pm \beta$ compared to the a site of the $a_s$ and the $a_{s+1}$ sites respectively.
}
\label{Fig-7}
\end{figure}

To observe the effect of sinusoidal driving, we fabricated five sinusoidally modulated rhombic lattices ($70$-mm-long) with the amplitude of modulation $A\!=\!6$~$\mu$m to 14~$\mu$m in steps of 2~$\mu$m, and the spatial frequency $\omega\!=\!1.5708$~mm$^{-1}$. The output intensity distributions for two types of input states are shown in Fig.~\ref{Fig-6}. It was observed that the flat-band modes are not affected by the sinusoidal driving whereas the output intensity distribution for equal-phase mode is determined by $K/\omega$.  Note that for Fig.~\ref{Fig-6}(g), the ratio of the field magnitude and the field frequency (i.e.~$K/\omega$) is close to the first root of $\mathcal{J}_0$ causing coherent destruction of tunneling.
The effective coupling constant of a modulated lattice, for a given value of $K$ and $\omega$ was evaluated by simulating a $70$-mm-long straight photonic lattice, and varying the coupling strength to optimally fit the observed output intensity distribution. The estimated values of normalized effective coupling constants, $|\kappa_{\text{eff}}/\kappa|$, are  in agreement with the theoretical prediction
as shown in Fig.~\ref{Fig-7}(a).

The robust localization of the flat-band modes in the rhombic chain can also be explained intuitively. In the presence of a static or ac driving, the effective propagation constant (or the analogous site energy) of the waveguides in the lattice at a given value of $z$ is determined by the driving strength and its direction~\cite{heiblum1975analysis, lenz1999bloch}. If the lattice is driven along the lattice axis, the $b$ and $c$ sites of the s-{th} unit cell have the same effective propagation constant which is shifted by $\pm\beta$ compared to the $a$ site of the $s$-{th} and the $(s+1)$-{th} unit cell respectively, see Fig.~\ref{Fig-7}(b). In this situation, the flat-band input state cannot tunnel to the nearest neighbor sites due to destructive interference as for the undriven case. 
Note that an extended initial flat-band state in real space will also exhibit this robust localization.
On the basis of this argument, one can infer that the symmetry in the lattice geometry is crucial to observe this robust localization in the presence of both static and ac drivings.

In conclusion, we investigated the dynamics of the initial states prepared in the dispersive and non-dispersive bands of the photonic rhombic lattices in the presence of both static and high-frequency sinusoidal drivings along the lattice axis. We experimentally excited flat-band eigenmodes and show that they remain localized in the presence of these drivings. However, when dispersive bands were excited by launching an orthogonal input state, we observed Bloch oscillations and coherent destruction of tunneling respectively. This robust flat-band localization is due to destructive interference of the analogous wavefunction and is associated with the symmetry in the lattice geometry. 

\vspace*{2 mm}

\noindent
\textbf{Data.$-$} Raw experimental data are available through Heriot-Watt University PURE research data management system~\cite{data}.


\noindent
\textbf{Acknowledgments.$-$} S. M. thanks Mr.~Alexander Spracklen (UK) and Prof.~Nathan Goldman (Belgium) for helpful comments. 

\vspace*{0.5 mm}

\noindent
\textbf{Funding.$-$} UK Science and Technology Facilities Council (STFC) ST/N000625/1.

\end{document}